\documentclass[%
 reprint,
 amsmath,amssymb,
 aps,
]{revtex4-2}
\usepackage{ragged2e}
\newtheorem{theorem}{Theorem}[section]
\usepackage{xcolor}
\usepackage{colortbl}
\usepackage{hyperref}
\usepackage{graphicx}
\usepackage{dcolumn}
\usepackage{bm}

\graphicspath{{figs/}}

\begin{document}

\preprint{APS/123-QED}

\title{Graph-based Complexity and Computational Capabilites of Proteinoid Spike Systems}


\author{Saksham Sharma}
\affiliation{Cambridge Centre for Physical Biology, CB3 0WA, Cambridge, UK}
\author{Adnan Mahmud}
\affiliation{Zuse Institute Berlin, Takustraße 7 14195, Germany}
\author{Giuseppe Tarabella}
\affiliation{Institute of Materials for Electronic and Magnetism, National Research Council
(IMEM-CNR), Parma (Italy)}
\author{Panagiotis Mougoyannis}
\author{Andrew Adamatzky}
\affiliation{
Unconventional Computing Laboratory, UWE Bristol, UK
}%
\date{\today}

\begin{abstract}
Proteinoids, as soft matter fluidic systems, are computational substrates that have been recently proposed for their analog computing capabilities. Such systems exhibit oscillatory electrical activity because of cationic and anionic exchange inside and outside such gels. It has also been recently shown that this (analog) electrical activity, when sampled at fixed time intervals, can be used to reveal their underlying information-theoretic, computational code. This code, for instance, can be expressed in the (digital) language of Boolean gates and QR codes. Though, this might seem as a good evidence that proteinoid substrates have computing abilities when subjected to analog-to-digital transition, the leap from their underlying computational code to computing abilities is not well demonstrated. In this work, we analyse the electrical activity patterns of proteinoid substrates using computational methods including deep ReLU networks. Our findings suggest intriguing parallels between these patterns and certain computational frameworks, hinting at the possibility that such chemical systems might possess inherent information-processing and computational capabilities. To demonstrate the computational ability, we construct a prediction algorithm which acts as a binary classification model and extract 16-dimensional vector data from the proteinoid spike, in order to perform predictions with 70.41\% accuracy. This model in its core has a unique transformation modality, inspired from number-theoretic sieve theory, and is combination of two functions: \textit{spiral sampling} $F_{1}$ and \textit{significant digit extraction} $F_{2}$ functions. The complexity of the transformed data is measured using eight distinct metrics, and effectively, using a single \textit{meta-metric}. \newline

\end{abstract}

\maketitle
Proteinoids, made from poly(amino) acids, exhibit oscillatory analog electrical activity because of the cationic and ionic exchanges inside their gelatinous structure. This analog information can be interpreted in a digital format, by conversion to Boolean gates and QR codes. Despite this possibility, it is not yet clear, as to what can make such systems capable of performing universal computation similar to how a biological neuron does. Though relatively young in the literature on proteinoids, there are many ``fluidic" soft matter systems that have been shown to have such neuron-like computing capabilities. In early 2000, Maass \textit{et al.} developed \textit{Liquid State Machine} consisting of a cortical microcolumn that connects the neurons randomly and is capable of universal real-time computation \cite{maass2002real}. This model was subsequently adopted ``literally" by considering water waves in a reservoir and training the system to solve XOR problem and perform speech recognition \cite{fernando2003pattern}. \newline 

There exists plenty of reservoir computing architectures in the literature, built out of fluidic systems upon implementation of neural algorithms. A few such fluidic computing systems include the systems for signal analysis and data classification operations \cite{maksymov2023reservoir}, computation at the nonlinear regime \cite{nakajima2020physical, marcucci2023new, marcucci2020theory}, and incorporating external force fields into the system such as acoustic fields \cite{mussel2019similarities, mussel2019sounds}. \newline

\begin{figure*}[!tbp]
  \includegraphics[width=1.0\textwidth]{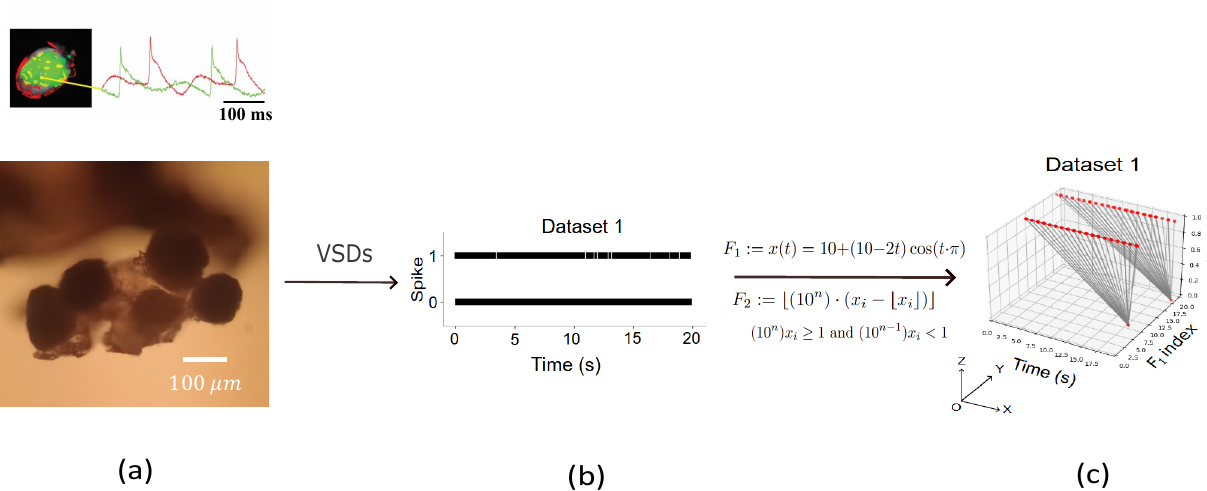}
  \caption{(a) Voltage versus time signal of the oscillatory electrical activity inside proteinoids measured using voltage sensitive dyes (top) and the microscopic images of the proteinoid microspheres (bottom). (b) Voltage sensitive dyes (VSDs) convert the recordings into a spike-versus-time data (Dataset 1 is plotted), (c) Functions $F_{1}$ and $F_{2}$ are used to transform Dataset 1 into a multi-nodal graph further detailed in Section \ref{sec:3}.}
   \label{fig:fig1}
\end{figure*}
 \begin{figure*}[!tbp]
  \includegraphics[width=1.0\textwidth]{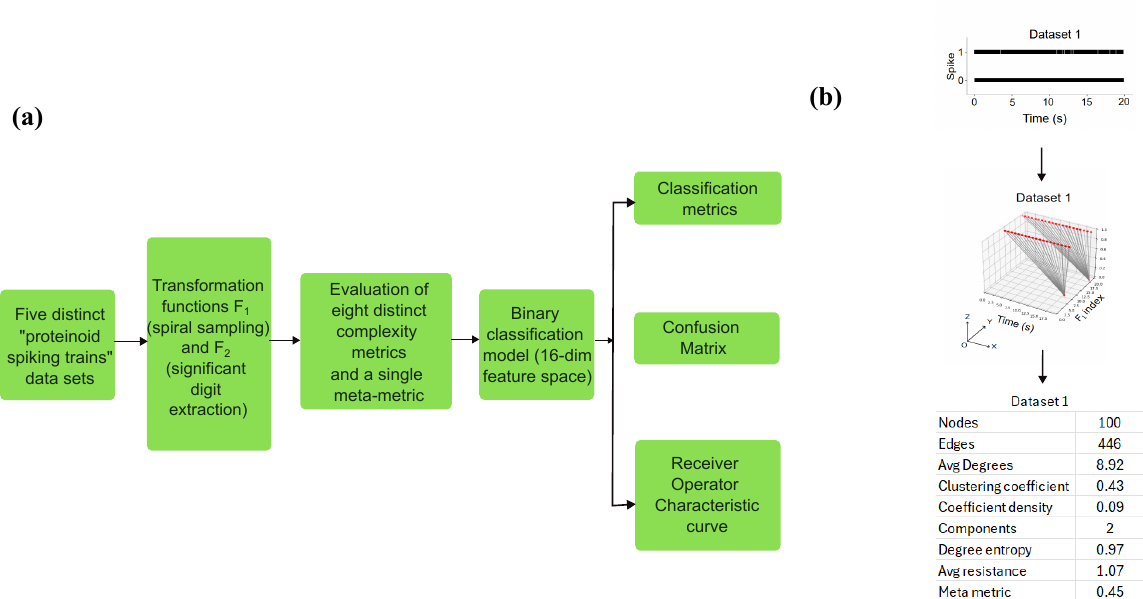}
  \caption{(a) The steps involved in the analysis shown in the current work are presented in a block fashion. (b) Data from Dataset 1 is transformed from spike-versus-time to a multi-nodal graph format with value of complexity metrics shown below.}
   \label{fig:fig1_5}
\end{figure*}
One crucial step in implementing such \textit{fluidic reservoir computing} algorithms is that the response of the physical system to an input impulse or signal is characterised and used for training a given neural network architecture. A plenty of such architectures exist in the literature, such as the Spiking Neural Networks (SNNs) \cite{severa2016spiking}, Extreme Learning Machines (ELMs) \cite{marcucci2020theory}, Echo State Networks (ESNs) \cite{heyder2021echo}, Recurrent Neural Network (RNN) \cite{siegelmann1995computation}, or more broadly, neuromorphic computing (NMC) architectures \cite{noy2023nanofluidic}, \cite{sharma2022navier}, \cite{sharma2022complexity}. While each of such networks have their own dictionary and methodology required to implement them, a generic fluidic system (proteinoid gel in our case), need a fundamental ``theory of computing (ToC)" framework that grants them inherent capability of universal computing, upon which a suitable NMC methodology can be implemented. \newline
\begin{figure*}
\centering
\includegraphics[width=\textwidth]{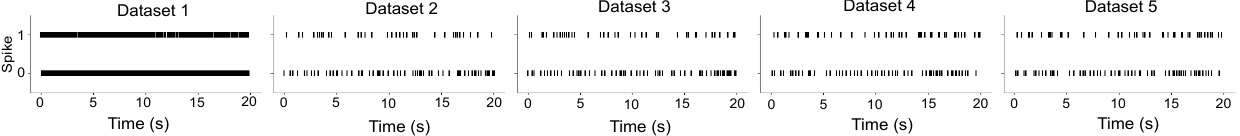}
\caption{Raw Spike Trains Datasets: The figure shows ``spikes or no spikes" (0s or 1s) in the proteinoid samples prepared in Section \ref{sec:protocol} plotted against time (s) and recorded using voltage sensitive dyes (VSDs).}
\label{fig:spike_trains}
\end{figure*}

In the current work, we record the electrical activity in proteinoid samples in the form of proteinoid spiking data. Five distinct datasets of the spiking trains extracted from the proteinoids are transformed using a novel modality, which is the combination of two functions: $F_{1}$ (spiral sampling) and $F_{2}$ (significant digit extraction). After this transformation, the complexity of the transformed data is calculated using eight graph-based discrete metrics which are then unified into a single \textit{meta-metric}. Finally, the prediction algorithm for the proteinoid spikes is constructed using a feedforward neural network architecture, which captures three class of features (temporal, statistical, and spectral) comprised of a 16-dimensional  vector space. The algorithm is, effectively,  classification model which works with 70.41\% accuracy (owing to a restricted data set from the VSD measurements). \newline

Broadly speaking, our analysis lets us motivate future directions for this research towards more universal (chemical) computing paradigms, such as proteinoid transformers or (advanced) proteinoid physical generative AI models, as opposite to primitive information-theoretic agents \cite{sharma2024information}, which can inspire experimentalists to come up with the physical embodiments of proteinoid GPTs - the collection of which can serve as the universal-computing multiverse with applications ranging from organics electronics technologies (OECTs) \cite{degiorgio2025}, microfluidics \& drug discovery \cite{pitingolo2018beyond}, to catalytic nanomotors \cite{dhar2020self} and beyond, and amenable for further relatively \textit{topical} investigations, such as the `c-word' of such physical systems.

\section{Protocol for measuring electrical activity inside proteinoids}
\label{sec:protocol}
The proteinoids are prepared using the protocol used before by us \cite{sharma2023morphological} and reviewed thoroughly here \cite{sharma2022review}. The experimental setup is described as the following: amino acids in powdered form; 3-hole round bottom flasks; Tri-block heater; $N_{2}$ gas; dialysis cellophane membrane; magnetic stirrer; and a water bath. 1.5 grams each of L-Aspartic acid, L-Histidine, L-Phenylalanine, L-Glutamic acid, and L-Lysine is mixed-and-heated in the (3-hole) round bottom flask at 290$^{\circ}$C. The temperature is step-by-step increased by 10$^{\circ}$. After fuming start to appear, the exhaust in the fumehood (with N$_{2}$ gas as the inlet) is started, to release the fumes out. The powder goes through a colour transformation from white to green colour, followed by a morphological transition resulting in the formation of a sequence of simmering microspheres. After ceasing the heating process, the remaining material solidifies and is subsequently extracted and allowed to cool for a duration of thirty minutes. The collected residue is then placed within the Slide-A-Lyzer mini dialysis apparatus, with 10,000 molecular weight cut-off and employs water as the dialysate. Dialysis is carried out continuously over for five days until the residue comprises microscopic ensembles of microspheres. The residue derived from the dialysis membrane is heated in the vacuum oven for half an hour, facilitating the evaporation of the dialysate (water) from the sample. Subsequently, the sample is scrutinised utilizing a transmission electron microscope. Voltage-sensitive (aminonaphthylethenylpyridinium) dyes are used for recording the electrical activity. During the recording process, the data logger (ADC-24, Pico Technology, UK) operated at its maximum capacity (600 data points per second). This rate of data collection allowed for a comprehensive understanding of the electrical activity exhibited by the proteinoids. Furthermore, the data logger stored and saved the average value obtained from these measurements. This approach provided a concise representation of the recorded electrical activity, facilitating subsequent analysis and interpretation of the experimental results.

\section{Chemical primordial soup to a univeral approximating network: a modified Kolmogorov-Arnold representation}

\subsection{Initial Data}

The initial dataset comprises of five distinct proteinoids spike trains, each representing the firing patterns of individual spikes against time. These spike trains are characterised by a series of temporal points, where each point signifies the timestamp of the firing event. The data is structured as follows:

\begin{enumerate}
    \item \textbf{Time series data:} Each spike train is represented as a time series. X-axis denotes time (in seconds) and Y-axis is binary (0 or 1), indicating the presence or absence of a spike.
    \item \textbf{Temporal resolution:} The timing of spikes is recorded with precision up to the order of microseconds ($\mu s$).
    \item \textbf{Variable duration:} Each spike train may have a different total duration, reflecting the variability in spiking activity periods.
    \item \textbf{Sparse nature:} Consistent with typical spiking firing patterns, the spike trains exhibit a sparse structure—--with relatively few spikes distributed over the recorded time period.
\end{enumerate}

\subsection{Transformation Process}

To extract deeper insights and reveal potential hidden structures within the spike train data, we apply a novel transformation process described ahead. The transformation converts the (sequential) time series data into a multi-nodal graph structure. \newline

Figure \ref{fig:spike_trains} displays the original spike train data for all five datasets. Each row represents a separate dataset, with time (in seconds) on the x-axis and spiking events represented by the vertical black lines on the y-axis. Dataset 1 exhibits a notably dense spike pattern, while Datasets 2-5 show varying degrees of sparsity and rhythmicity in their spiking patterns. Temporal clustering of spikes is evident in some datasets, particularly in Datasets 3-5.\newline

The transformation process involves two key functions, F1 and F2, which operate as follows:
\subsubsection{Function $F_{1}$: Spiral Sampling}
\noindent 
\begin{equation}
F_{1}:= x(t) = 10 + (10 - 2t) \cos(t \cdot \pi), \quad t \in \{0, 2, 3, ..., 19, 20\} 
\end{equation}

\noindent $F_{1}$ is designed to perform non-uniform sampling mechanism and select points from the given spike train in a spiraling inward fashion. As $t$ ranges from 0 to 20, $x(t)$ generates a series of values that oscillate between 0.5 and 20, with an inward trend. For example, $x(13)$ gives 6.5, and the points in the dataset \textit{just less} than 6.5 are picked. \newline

\noindent The key aspects of F1 are:

\begin{enumerate}
    \item \textbf{Non-linear sampling:} Unlike uniform sampling, $F_{1}$ provides a non-linear distribution of sampling points, potentially revealing patterns at different temporal scales.
    \item \textbf{Bounded output:} The output of the function is bounded between 0.5 and 20, ensuring that the sampling remains within a predefined range regardless of the input spike train's duration.
    \item \textbf{Decreasing periodicity:} The cosine term introduces an oscillatory behavior with decreasing amplitude, mimicking a spiral pattern when visualised.
\end{enumerate}

\subsubsection{Function $F_{2}$: Significant Digit Extraction}

\begin{equation}
F_{2} := \left\lfloor (10^n) \cdot (x_i - \left\lfloor x_i \right\rfloor) \right\rfloor; (10^n)x_i \geq 1 \text{ and } (10^{n-1})x_i < 1
\end{equation}
where $x_i$ represents a time point from the original spike train.\newline

This function extracts the first significant digit after the decimal point for each selected time point. The key aspects of $F_{2}$ are:

\begin{enumerate}
    \item \textbf{Scale invariance:} By focusing on the first significant digit, $F_{2}$ captures a scale-invariant property of the time points, potentially revealing patterns that are independent of the absolute time scale.
    \item \textbf{Digit-based clustering:} Points with the same first significant digit are grouped together, creating a natural clustering mechanism based on this mathematical property. The reference point would constitute the hypothetical data point which end with ``.000".
\end{enumerate}

\section{Transformation Procedure \& Effects} \label{sec:3}

\begin{figure*}
\centering
\includegraphics[width=\textwidth]{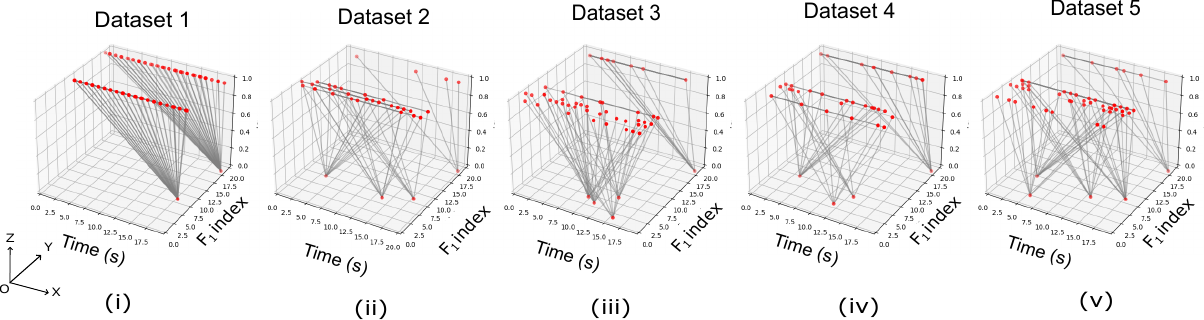}
\caption{Multinodal Graph Transformation spike trains datasets: subplots (i-v) represent a transformed dataset, where the x-axis represents time in seconds, the y-axis represents the $F_{1}$ index, and the z-axis (vertical) represents the connection layer. Red nodes indicate the primary nodes selected by the $F_{1}$ function and the grey edges connect the primary nodes to their corresponding secondary nodes.}
\label{fig:graph_transformation}
\end{figure*}

Figure \ref{fig:graph_transformation} illustrates the outcome of applying the transformation process to Datasets 1-5, which involves converting the linear spike trains into complex, three-dimensional graph structures. Table \ref{alg:transformation}
outlines six algorithmic steps of the transformation. \newline

\begin{table*}    
\caption{Data Transformation Algorithm}
\label{alg:transformation}
\renewcommand{\arraystretch}{1.3}
\begin{tabular}{p{\textwidth}}
\rowcolor[gray]{0.9}
\textbf{Input:} Original spike train dataset $S$, Functions $F_{1}$ and $F_{2}$ \\
\rowcolor[gray]{0.9}
\textbf{Output:} Graph $G$ representing analyzed spike train \\
\rowcolor[gray]{0.9}
\textbf{Algorithm:} \\
\rowcolor[gray]{0.9}
1. Initialise empty sets: \\
\rowcolor[gray]{0.9}
\quad $P \gets \emptyset$ (Set of primary nodes) \\
\rowcolor[gray]{0.9}
\quad $D \gets \emptyset$ (Set of secondary nodes) \\
\rowcolor[gray]{0.9}
\quad $E \gets \emptyset$ (Set of edges) \\
\rowcolor[gray]{0.9}
2. For each $t \in \{0, 2, ..., 20\}$: \\
\rowcolor[gray]{0.9}
\quad a) Compute $x(t) \gets F_{1}(t)$ \\
\rowcolor[gray]{0.9}
\quad b) Find $p = \max\{s \in S : s \leq x(t)\}$ (closest point not exceeding $x(t)$) \\
\rowcolor[gray]{0.9}
\quad c) Add to primary nodes: $P \gets P \cup \{p\}$ \\
\rowcolor[gray]{0.9}
3. For each $p \in P$: \\
\rowcolor[gray]{0.9}
\quad Compute $d_p \gets F_{2}(p)$ (Extract first significant digit after decimal) \\
\rowcolor[gray]{0.9}
4. For each $s \in S$: \\
\rowcolor[gray]{0.9}
\quad a) Compute $d_s \gets F_{2}(s)$ \\
\rowcolor[gray]{0.9}
\quad b) For each $p \in P$: \\
\rowcolor[gray]{0.9}
\quad\quad If $d_s = d_p$: \\
\rowcolor[gray]{0.9}
\quad\quad\quad Add to secondary nodes: $D \gets D \cup \{s\}$ \\
\rowcolor[gray]{0.9}
\quad\quad\quad Add edge: $E \gets E \cup \{(p, s)\}$ \\
\rowcolor[gray]{0.9}
5. Construct graph: $G \gets (P \cup D, E)$ \\
\rowcolor[gray]{0.9}
6. Return $G$ \\
\end{tabular}
\end{table*}

The transformation acts on the representation of the spike train data and causes the following changes:

\begin{enumerate}
    \item \textbf{Dimensionality Increase:} The original 1-dimensional time series is transformed into a 3-dimensional structure (time, $F_{1}$ index, and connection layer).
    \item \textbf{Topology Change:} The linear structure of the time series is converted into a complex graph with multiple interconnected nodes.
    \item \textbf{Multi-scale Representation:} The combination of $F_{1}$ spiral sampling and $F_{2}$ digit extraction creates a multi-scale representation, with an aim to reveal both large-scale temporal patterns and fine-grained similarities between the spike timings.
    \item \textbf{Pattern Emergence:} The graph structure may reveal clusters or patterns of spike timings that were not readily apparent in the original linear representation.
    \item \textbf{Information Condensation:} While some temporal information is lost in the transformation, the resulting structure condenses information about similarities and patterns in spike timing across different time scales.
\end{enumerate}

\section{Complexity Metrics and Meta-Metric evaluation}\label{sec:4}

\subsection{Discrete Complexity Metrics}

To quantify the complexity of the transformed spike train data, as noted in Table \ref{tab:complexity_metrics_1} and \ref{tab:complexity_metrics_2}, we evaluate a set of graph-based metrics. These metrics capture various aspects of the multinodal graph structure, providing insights into the intricacy and patterns within the neuronal firing data. We then combine these metrics into a \textit{meta-metric} to obtain an overall measure of complexity.

\begin{figure*}
\centering
\includegraphics[width=0.7\textwidth]{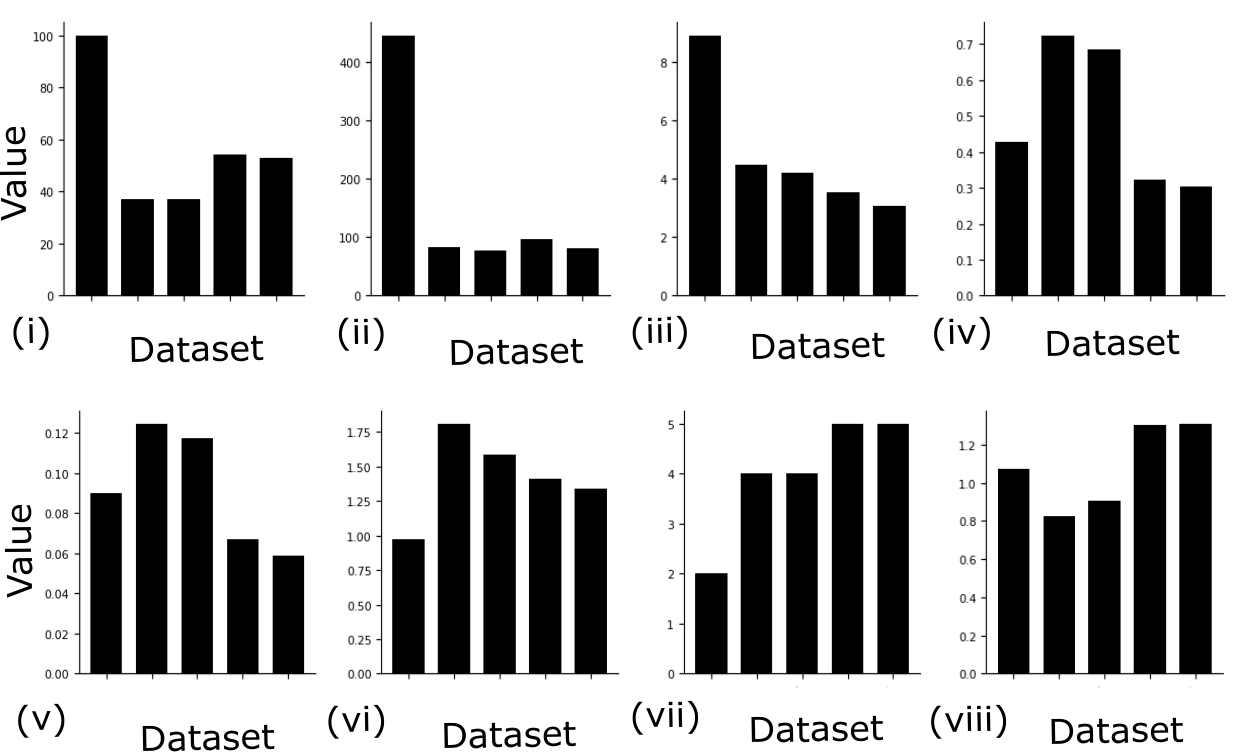}
\caption{Comparison of Complexity Metrics Across Datasets (in linear order: 1, 2, 4, 5, and 3). Y-axis data correspond to: (i) number of nodes $(N)$, (ii) number of edges $(E)$, (iii) average degree $(\bar{k})$, (iv) clustering coefficient $C$, (v) graph density $\rho$, (vi) degree entropy $H$, (vii) number of connected components $CC$, (viii) average resistance $R$.}
\label{fig:complexity_metrics}
\end{figure*}

\begin{table*}
\caption{Discrete Complexity Metrics}
\label{tab:complexity_metrics_1}
\resizebox{\textwidth}{!}{%
\begin{tabular}{p{0.2\textwidth}p{0.4\textwidth}p{0.4\textwidth}}
\hline
\multicolumn{1}{c}{\textbf{Metric}} & 
\multicolumn{1}{c}{\textbf{Description}} & 
\multicolumn{1}{c}{\textbf{Equation}} 

\\
\hline
Number of Nodes ($N$) & Counts the total number of nodes in the graph. &  $N = |V|$, where $V$ is the set of vertices in the graph \\
Number of Edges ($E$) & Counts the total number of connections in the graph. & $E = |E|$, where $E$ is the set of edges in the graph \\
Average Degree ($\bar{k}$) & Measures the average number of connections per node. & $\bar{k} = \frac{2E}{N}$ \\
Clustering Coefficient ($C$) & Quantifies the degree to which nodes in the graph tend to cluster together. & $C = \frac{1}{N} \sum_{i=1}^N C_i$, where $C_i$ is the local clustering coefficient of node $i$ \\
Graph Density ($\rho$) & Measures how close the graph is to being complete. & $\rho = \frac{2E}{N(N-1)}$ \\
Degree Entropy ($H$) & Quantifies the complexity of the degree distribution. & $H = -\sum_{k} P(k) \log P(k)$, where $P(k)$ is the probability of a node having degree $k$ \\
Number of Connected Components ($CC$) & Counts the number of disconnected subgraphs within the overall graph structure. & N/A \\
Average Resistance ($R$) & Measures the overall connectivity structure within components. & $R = \frac{1}{|CC|} \sum_{c \in CC} \frac{1}{\binom{|V_c|}{2}} \sum_{i<j \in V_c} (L^+_{ii} + L^+_{jj} - 2L^+_{ij})$, where $L^+$ is the pseudoinverse of the Laplacian matrix, and $V_c$ is the set of vertices in component $c$ \\
\hline
\end{tabular}
}
\end{table*}

\subsection{Meta-Metric for Overall Complexity}

To combine these discrete metrics into a single measure of overall complexity, let us define the \textit{meta-metric} as the metric that provides a nuanced ranking of complexity and avoids artificial bounds and reflects the continuous nature of complexity. It is calculated as follows:

\begin{enumerate}
\item \textbf{Normalisation:} Each metric is normalised using the notion of z-score and to ensure comparability across different scales:
\begin{equation}
z_i = \frac{x_i - \mu_i}{\sigma_i}
\end{equation}
where $x_i$ is the original metric value, $\mu_i$ is the mean, and $\sigma_i$ is the standard deviation of metric $i$ across all datasets.
\item \textbf{Weighted Sum:} A weighted sum of the normalised metrics is computed
\begin{equation}
S = \sum_{i=1}^8 w_i z_i
\end{equation}
where $w_i$ are the weights assigned to each metric \footnote{The weights $w_i$ are as follows: Number of Nodes: 0.05, Number of Edges: 0.05, Average Degree: 0.10, Clustering Coefficient: 0.10, Graph Density: 0.20, Degree Entropy: 0.20, Number of Connected Components: 0.10, and Average Resistance: 0.20. These weights were chosen to balance the contribution of each metric, with slightly higher emphasis on degree entropy and average resistance as they capture more nuanced aspects of the graph structure.}, reflecting their relative importance in determining overall complexity.
\item \textbf{Sigmoid Transformation:} The weighted sum is transformed using a sigmoid function:
\begin{equation}
\text{Meta-metric} = \frac{1}{1 + e^{-S}}
\end{equation}
\end{enumerate}

\noindent The sigmoid transformation ensures that the meta-metric asymptotically approaches 0 for absolute simple systems and 1 for absolute complex systems, without exactly ever reaching these values. This reflects the idea that complexity exists on a continuous spectrum, and allows for meaningful comparisons between datasets while leaving room for potentially more or less complex datasets in future analyses.

\subsection{Analysis of Complexity Metrics}

\noindent To analyse the complexity of the transformed spike train data across our five datasets, we compute eight individual metrics and a meta-metric for each dataset. Table \ref{tab:complexity_metrics_2} and Figure \ref{fig:complexity_metrics} provides a comprehensive comparison of these metrics across datasets 1-5. \newline

\begin{table*}
\centering
\caption{Comprehensive Complexity Metrics Comparison}
\label{tab:complexity_metrics_2}
\resizebox{\textwidth}{!}{%
\begin{tabular}{ccccccccccc}
\hline
\textbf{Rank} & \textbf{Dataset} & \textbf{Nodes} & \textbf{Edges} & \textbf{Avg Degree} & \textbf{Clustering Coef.} & \textbf{Density} & \textbf{Components} & \textbf{Degree Entropy} & \textbf{Avg Resistance} & \textbf{Meta Metric} \\
\hline
1 & Dataset 2 & 37 & 83 & 4.4865 & 0.7240 & 0.1246 & 4 & 1.8104 & 0.8273 & 0.5789 \\
2 & Dataset 4 & 37 & 78 & 4.2162 & 0.6855 & 0.1171 & 4 & 1.5879 & 0.9039 & 0.5353 \\
3 & Dataset 5 & 54 & 96 & 3.5556 & 0.3215 & 0.0671 & 5 & 1.4095 & 1.3010 & 0.4826 \\
4 & Dataset 1 & 100 & 446 & 8.9200 & 0.4278 & 0.0901 & 2 & 0.9759 & 1.0745 & 0.4568 \\
5 & Dataset 3 & 53 & 81 & 3.0566 & 0.3040 & 0.0588 & 5 & 1.3389 & 1.3095 & 0.4461 \\
\hline
\end{tabular}%
}
\end{table*}

Complexity is not solely determined by the graph size. Dataset 1 ranks fourth despite being the largest, partly because of local connectivity and that the clustering coefficient is a strong contributor to overall complexity. Dataset 2 attains the highest complexity because it contains a fine balance of the complexity metrics. In addition, connected components play a nuanced role in measurement of the complexity. Additionally, degree entropy is a critical factor, in that it suggests the importance of diverse node connections. High complexity of the Dataset 2 represents rich and structured spiking interactions, whereas, Dataset 1 represents high-and-uniform activity, and a balance between local clustering and overall connectivity which further indicates efficient information processing. Multiple connected components in Datasets 3 and 5 suggest distinct sub-networks. \newline

The importance of individual metrics varies, depending on specific spiking phenomena and a further analysis would be needed on the temporal evolution of complexity measures. There is also a possibility of correlating complexity measures with specific spiking functions or stimuli responses -- therefore, a tailored analysis focusing on specific aspects of spiking dynamics or comparative studies is necessary. \newline


\section{Classification Model for Spike Train Analysis}\label{sec:5}

We developed a binary classification model aimed at predicting proteinoid spikes, in order to extract insights into the temporal dynamics of the firing.  

\subsection{Model Architecture}

We implemented a feedforward neural network using dense layers for our classification task. The architecture of our model is as follows:

\begin{itemize}
    \item \textbf{Input Layer}: It corresponds to the number of engineered features detailed in Section \ref{subsec:feature_engineering}.
    \item \textbf{Hidden Layers}: Four dense layers with 128, 64, 32, and 16 neurons respectively.
    \item \textbf{Output Layer}: A single neuron with sigmoid activation for binary classification.
\end{itemize}

All hidden layers utilise the Rectified Linear Unit (ReLU) activation function, which introduces non-linearity and helps mitigate the vanishing gradient problem during training.

\subsection{Feature Engineering}
\label{subsec:feature_engineering}

\begin{figure*}
\centering
\includegraphics[width=1.0\textwidth]{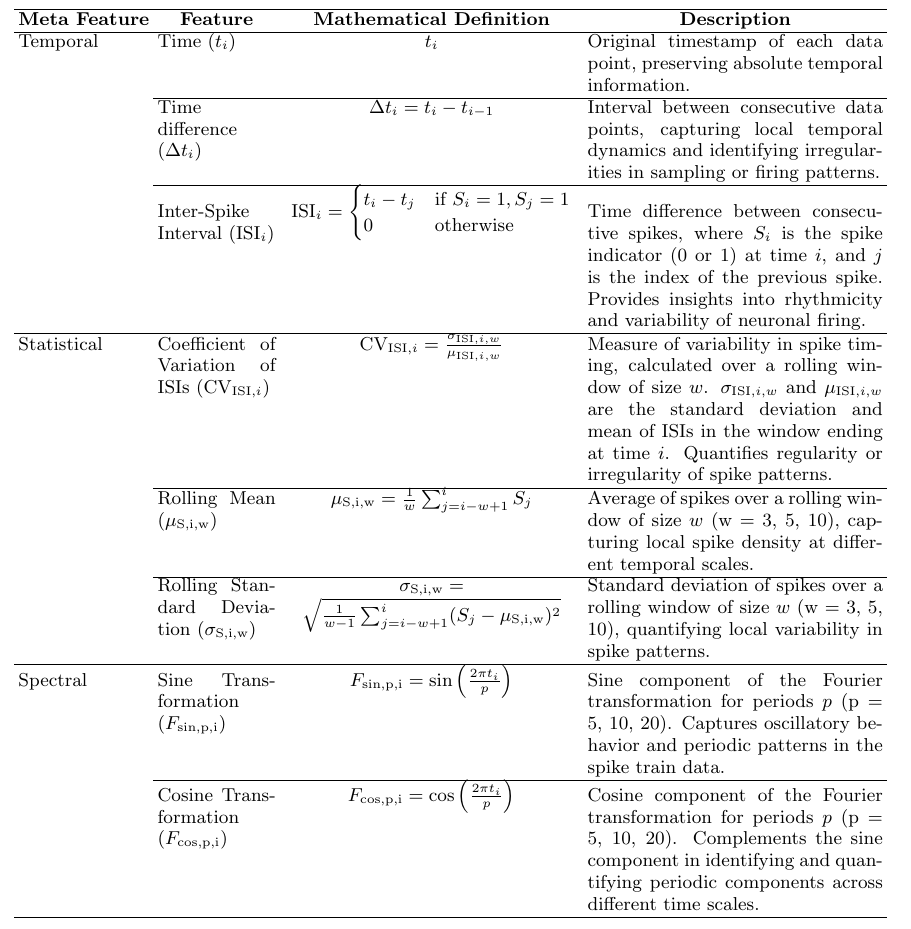}
\caption{Comprehensive Feature Set for Spike Train Classification}
\label{tab:all_features}
\end{figure*}

Table \ref{tab:all_features} presents a comprehensive and technical overview of all features used in our spike train classification model. These features are designed to capture complex temporal dynamics and structural characteristics of the neuronal activity. \newline

\noindent The complete feature vector $\mathbf{x}_i$ for each time point $i$ is constructed as follows

\begin{equation}
\begin{aligned}
\mathbf{x}_i = [&t_i, \Delta t_i, \textrm{ISI}_i, \textrm{CV}_{\textrm{ISI},i}, \mu_{S,i,3}, \sigma_{S,i,3}, \\
&\mu_{S,i,5}, \sigma_{S,i,5}, \mu_{S,i,10}, \sigma_{S,i,10}, \\
&F_{sin,5,i}, \textrm{F}_{cos,5,i}, \textrm{F}_{sin,10,i}, \textrm{F}_{cos,10,i}, \\
&\textrm{F}_{sin,20,i}, \textrm{F}_{cos,20,i}]
\end{aligned}
\end{equation}

This 16-dimensional feature vector provides a rich representation of the spike train data, encompassing temporal, statistical, and spectral characteristics. By leveraging this comprehensive feature set, our classification model captures intricate patterns in the spiking firing behavior and enables accurate spike prediction in the proto-cognitive agent data.

\subsection{Model Performance Analysis}

\subsubsection{Classification Metrics}

\begin{table}[h]
\centering
\begin{tabular}{lr}
\hline
\multicolumn{1}{c}{\textbf{Metric}} & 
\multicolumn{1}{c}{\textbf{Value}}\\
\hline
True Positives & 24 \\
False Positives & 10 \\
True Negatives & 45 \\
False Negatives & 19 \\
Accuracy & 0.7041 \\
Precision & 0.7059 \\
Recall & 0.5581 \\
F1 Score & 0.6234 \\
\hline
\end{tabular}
\caption{Classification Performance Metrics}
\label{table:metrics}
\end{table}

As tabulated in Table \ref{table:metrics}, the model achieves an accuracy of 70.41\%, i.e., correctly classifying about 7 out of 10 instances. While this indicates better-than-chance performance, there is a room for improvement. With a precision of 70.59\%, the model demonstrates a good ability to avoid false positives. When the model predicts a spike, it is correct approximately 71\% of the time. The recall of 55.81\% suggests that the model identifies about 56\% of all actual spikes. This relatively lower recall indicates that the model misses a significant portion of true spikes. The F1 score of 0.6234 provides a balanced measure of the model's performance, considering both precision and recall. This score indicates moderate performance but also highlights the potential for enhancement.

\subsubsection{Receiver Operating Characteristic (ROC) Curve}

\begin{figure}[h]
\centering
\includegraphics[width=0.5\textwidth]{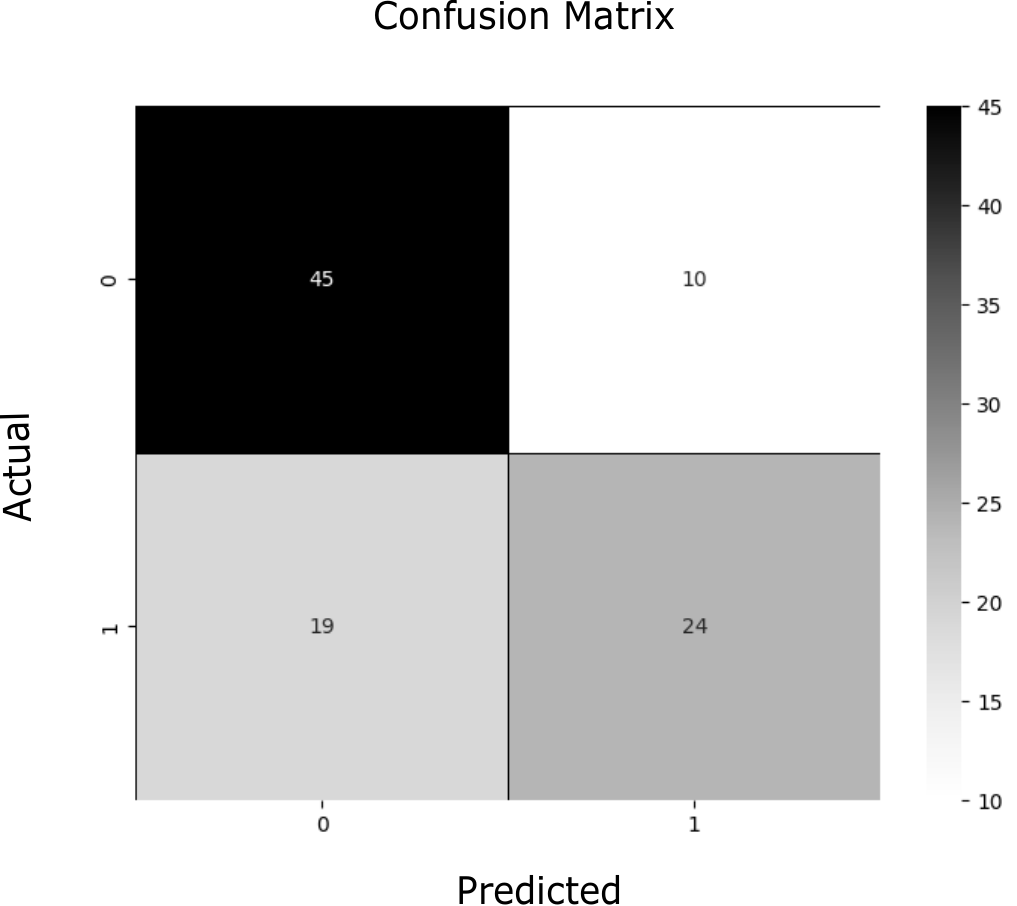}
\caption{Confusion Matrix (actual versus predicted) of the Spike Prediction Model.}
\label{fig:confusion_matrix}
\end{figure}
\begin{figure}
\centering
\includegraphics[width=0.5\textwidth]{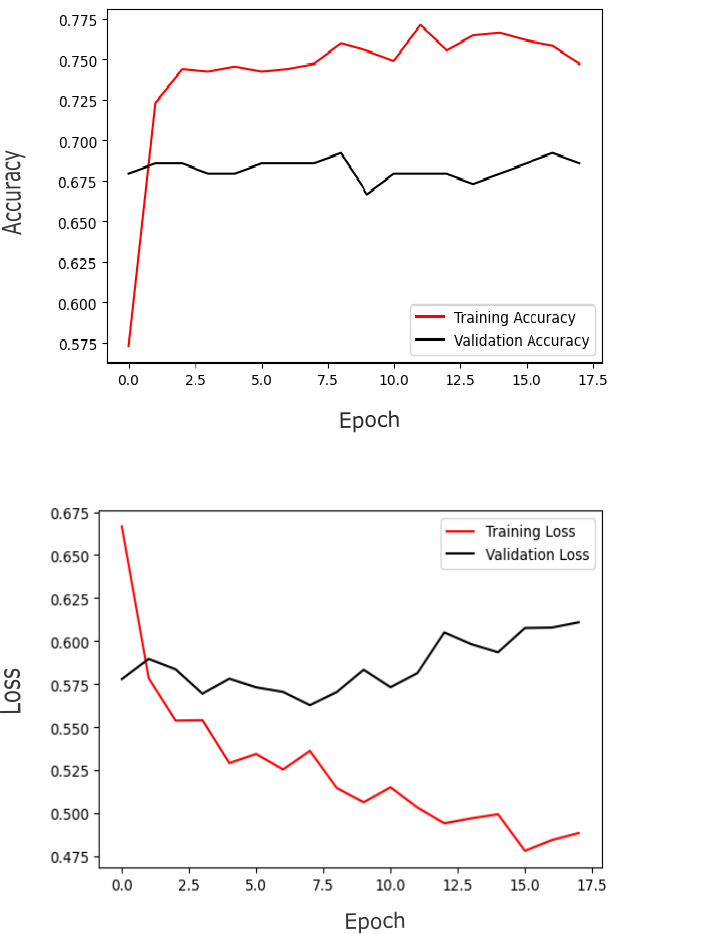}
\caption{Training and Validation Accuracy and Loss}
\label{fig:training_validation}
\end{figure}
\begin{figure}[h]
\centering
\includegraphics[width=0.5\textwidth]{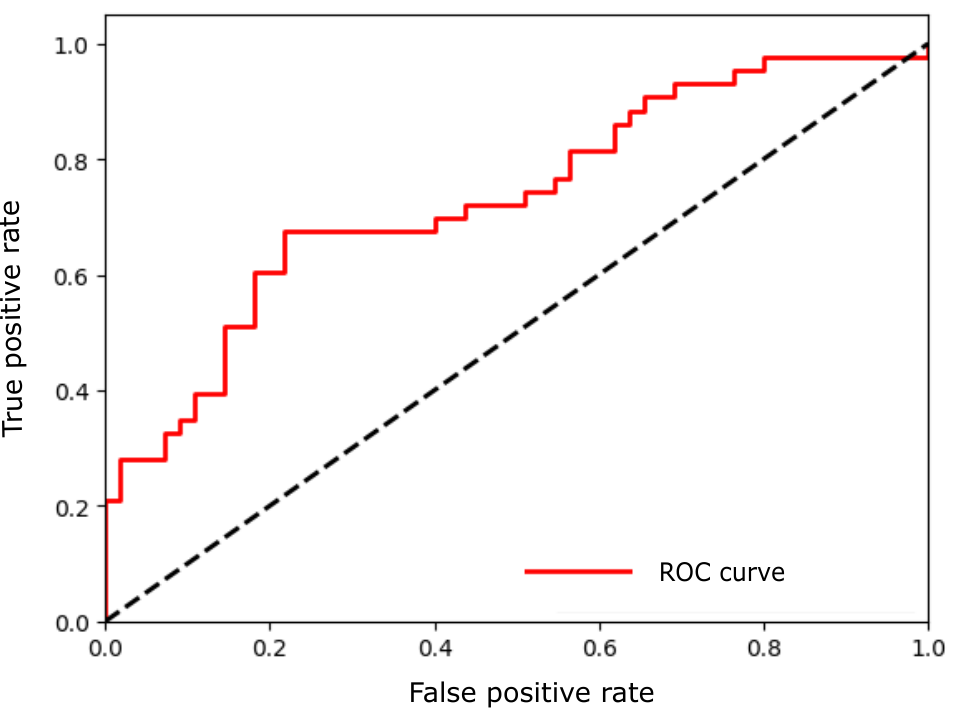}
\caption{Receiver Operating Characteristic (ROC) Curve with AUC = 0.73.}
\label{fig:roc_curve}
\end{figure}

Figure \ref{fig:roc_curve} illustrates the trade-off between the true positive rate (sensitivity) and the false positive rate (1 - specificity) at various classification thresholds. Our model achieves an Area Under the Curve (AUC) of 0.73, indicating a moderate discriminative ability. This AUC value suggests that the model performs better than random guessing (AUC = 0.5) in distinguishing between spike and non-spike events. \newline

The curve's shape reveals that our model achieves a relatively high true positive rate at lower false positive rates, as evidenced by the steep initial rise. This characteristic is desirable, especially in the context of neuronal spike detection where minimising false positives while maintaining sensitivity is crucial.

\subsubsection{Interpretation and Implications}

The model's performance metrics reveal several key insights:

1. \textbf{Balanced Performance:} The similar values of precision and accuracy suggest a relatively balanced performance, which is crucial in spike train analysis where both false positives and false negatives can significantly impact interpretations.

2. \textbf{Conservative Predictions:} The higher precision compared to recall indicates that the model is somewhat conservative in its spike predictions. It prioritizes avoiding false positives over capturing all spikes.

3. \textbf{Missed Spikes:} The lower recall value suggests that the model is missing a considerable number of actual spikes. This could lead to underestimation of neuronal activity in certain analyses.

4. \textbf{Potential for Improvement:} The moderate F1 score and AUC indicate that while the model performs better than random guessing, there is substantial room for improvement. This could potentially be achieved through feature engineering, model architecture refinement, or increased training data.

5. \textbf{Context-Dependent Utility:} The model's current performance may be more suitable for applications where minimizing false positives is prioritized over capturing every spike. However, for studies requiring high sensitivity to neuronal activity, further optimization would be beneficial.

In the context of analyzing protocognitive agents, these results suggest that our model can, potentially, provide valuable insights into spike patterns, but---of course---caution should be exercised when making definitive claims about neuronal activity based solely on these predictions. For example, it is worth noting that our application of neural networks to analyse proteinoid spike patterns does not, in itself, establish that proteinoids function as neural networks. Nevertheless, the effectiveness of these methods in extracting meaningful patterns suggests that proteinoid electrical activity contains rich information content with
potential computational relevance. The parallels observed raise intriguing questions about whether proteinoids might possess inherent computational properties that future research could more directly investigate through further targeted experimental approaches. \newline

\section{POTENTIAL RELATIONSHIP BETWEEN PROTEINOID ACTIVITY AND COMPUTATIONAL FRAMEWORK}
Our computational framework utilises neural networks with ReLU activation functions to process and get trained on proteinoid spike data. Intriguingly, this framework bears notable mathematical similarities to the modified Kolmogorov-Arnold representation theorem described by Schmidt-Heiber \cite{schmidt2021kolmogorov}. These similarities raise the possibility that
proteinoid dynamics might share certain computational properties with ReLU networks, though establishing such a connection would require devoted, further research. \newline

To hint further on this similarity, let us consider the network architecture as consisting of an input layer, output layer, and multiple hidden layers, where each hidden layer extracts bits from the binary representation of inputs to emulate the KA decomposition described in \cite{schmidt2021kolmogorov}. This process allows the network to approximate a target function efficiently with fewer parameters by computing intermediate bit-extraction operations and combining them to form a piecewise constant function. Through the use of ReLU activation functions, the network can closely simulate thresholding behaviors, which are essential for the KA structure. The theorem proved as a result is the following in \cite{schmidt2021kolmogorov}
\begin{theorem}
    Let \( p \in [1, \infty) \). If there exist \( \beta \leq 1 \) and a constant \( Q \), such that \( |f(\mathbf{x}) - f(\mathbf{y})| \leq Q |\mathbf{x} - \mathbf{y}|_{\infty}^{\beta} \) for all \( \mathbf{x}, \mathbf{y} \in [0,1]^d \), then, there exists a deep \textrm{ReLU} network \( \tilde{f} \) with \( 2K + 3 \) hidden layers, network architecture \( (2K+3, (d, 4d, \ldots, 4d, d, 1, 2^{Kd}+1, 1)) \) and all network weights bounded in absolute value by \( 2(Kd \vee \|f\|_{\infty})^{2K(d \vee (\beta p))} \), such that
\[
\|f - \tilde{f}\|_p \leq 2 \left( Q + \|f\|_{\infty} \right) 2^{-\beta K}.\]
\end{theorem}

\noindent Further details of this are given in Section 3 of \cite{schmidt2021kolmogorov}. We invite the interested readers to work on an exercise problem, of finding a ``mapping" of the parameters in the theorem above, to the typical computational routine followed by us (Section \ref{sec:3}-\ref{sec:5}) in training neural networks on the proteinoid spiking data.

\section{Conclusion and future directions}
In the present work, we are able to construct a \textit{modality} that converts the discrete time-series data using the recordings from proteinoid ensembles (using voltage sensitive dyes) to a multi-nodal graph structure, using two functions $F_{1}$ and $F_{2}$, which perform non-linear sampling from the given data and then extract the first significant digit after the decimal point from the datapoint. The data transformation algorithm employed in this process is unique to this analysis, and hope will be further used if applicable to other time-series data samples. We were largely motivated to characterise the data samples in terms of complexity metrics (conventional discrete complexity metrics and a novel \textit{meta-metric}). The metrics measured, as a result, helped us to compare different datasets and identify their individual nuances (clustering coefficient, local connectivity, sub-networks, etc). Finally, the binary classification model serves as a prediction tool to identify 16 key features from the spike train data. The tool works with 70.59\% accuracy which needs further improvement. We conclude by noting both the efficacy of ReLU networks as computational tools for
proteinoid spike data and the intriguing mathematical parallels observed. These parallels hint
at possible deeper relationships between proteinoid systems and established computational
frameworks such as the modified Kolmogorov-Arnold representation theorem \cite{schmidt2021kolmogorov}, opening avenues for further theoretical and experimental exploration. \newline 

We plan to further investigate the spiking train data from the proteinoid ensembles in our future studies. One potential direction that inspires us is that of asking: what \textit{fundamentally} differentiates the given dataset from a randomly generated sequence? While we know that the binary classification model can make such distinction, the underlying reason is not hinted yet, and will be the topic of our forthcoming study.  

\appendix

\section{Appendix}

This appendix provides additional visualisations of our spike train classification model's performance, offering deeper insights into its behavior during training and its predictive accuracy. Figure \ref{fig:confusion_matrix} presents the confusion matrix of our model's predictions on the test set. Figure \ref{fig:training_validation} shows the evolution of accuracy and loss for both training and validation sets over the course of model training. These plots provide several key insights:

\begin{itemize}
    \item \textbf{Accuracy (left plot):} The training accuracy (red line) shows a rapid increase in the early epochs, followed by a more gradual improvement. The validation accuracy (black line) remains relatively stable, suggesting that the model generalizes well to unseen data.
    
    \item \textbf{Loss (right plot):} The training loss (red line) decreases steadily throughout the training process, indicating continuous improvement in the model's performance on the training data. The validation loss (black line) shows some fluctuations but generally remains stable, further supporting the model's generalization capability.
    
    \item \textbf{Overfitting assessment:} The gap between training and validation metrics is relatively small, suggesting that the model is not severely overfitting to the training data.
\end{itemize}

These visualisations complement the quantitative metrics presented in the main text, providing a more comprehensive view of the model's learning process and predictive performance.

\section*{Code}
The computational framework and associated algorithms for proteinoid spike analysis are available in the Github repository \href{https://github.com/Adnan1729/Protocognitive_Agents/tree/main/main}{\textcolor{red}{here}}, with implementation by A.M.

\bibliography{proteinoidbib1}

\providecommand{\noopsort}[1]{}\providecommand{\singleletter}[1]{#1}%
\begin{thebibliography}{22}%
\makeatletter
\providecommand \@ifxundefined [1]{%
 \@ifx{#1\undefined}
}%
\providecommand \@ifnum [1]{%
 \ifnum #1\expandafter \@firstoftwo
 \else \expandafter \@secondoftwo
 \fi
}%
\providecommand \@ifx [1]{%
 \ifx #1\expandafter \@firstoftwo
 \else \expandafter \@secondoftwo
 \fi
}%
\providecommand \natexlab [1]{#1}%
\providecommand \enquote  [1]{``#1''}%
\providecommand \bibnamefont  [1]{#1}%
\providecommand \bibfnamefont [1]{#1}%
\providecommand \citenamefont [1]{#1}%
\providecommand \href@noop [0]{\@secondoftwo}%
\providecommand \href [0]{\begingroup \@sanitize@url \@href}%
\providecommand \@href[1]{\@@startlink{#1}\@@href}%
\providecommand \@@href[1]{\endgroup#1\@@endlink}%
\providecommand \@sanitize@url [0]{\catcode `\\12\catcode `\$12\catcode `\&12\catcode `\#12\catcode `\^12\catcode `\_12\catcode `\%12\relax}%
\providecommand \@@startlink[1]{}%
\providecommand \@@endlink[0]{}%
\providecommand \url  [0]{\begingroup\@sanitize@url \@url }%
\providecommand \@url [1]{\endgroup\@href {#1}{\urlprefix }}%
\providecommand \urlprefix  [0]{URL }%
\providecommand \Eprint [0]{\href }%
\providecommand \doibase [0]{https://doi.org/}%
\providecommand \selectlanguage [0]{\@gobble}%
\providecommand \bibinfo  [0]{\@secondoftwo}%
\providecommand \bibfield  [0]{\@secondoftwo}%
\providecommand \translation [1]{[#1]}%
\providecommand \BibitemOpen [0]{}%
\providecommand \bibitemStop [0]{}%
\providecommand \bibitemNoStop [0]{.\EOS\space}%
\providecommand \EOS [0]{\spacefactor3000\relax}%
\providecommand \BibitemShut  [1]{\csname bibitem#1\endcsname}%
\let\auto@bib@innerbib\@empty
\bibitem [{\citenamefont {Maass}\ \emph {et~al.}(2002)\citenamefont {Maass}, \citenamefont {Natschl{\"a}ger},\ and\ \citenamefont {Markram}}]{maass2002real}%
  \BibitemOpen
  \bibfield  {author} {\bibinfo {author} {\bibfnamefont {W.}~\bibnamefont {Maass}}, \bibinfo {author} {\bibfnamefont {T.}~\bibnamefont {Natschl{\"a}ger}},\ and\ \bibinfo {author} {\bibfnamefont {H.}~\bibnamefont {Markram}},\ }\bibfield  {title} {\bibinfo {title} {Real-time computing without stable states: A new framework for neural computation based on perturbations},\ }\href@noop {} {\bibfield  {journal} {\bibinfo  {journal} {Neural computation}\ }\textbf {\bibinfo {volume} {14}},\ \bibinfo {pages} {2531} (\bibinfo {year} {2002})}\BibitemShut {NoStop}%
\bibitem [{\citenamefont {Fernando}\ and\ \citenamefont {Sojakka}(2003)}]{fernando2003pattern}%
  \BibitemOpen
  \bibfield  {author} {\bibinfo {author} {\bibfnamefont {C.}~\bibnamefont {Fernando}}\ and\ \bibinfo {author} {\bibfnamefont {S.}~\bibnamefont {Sojakka}},\ }\bibfield  {title} {\bibinfo {title} {Pattern recognition in a bucket},\ }in\ \href@noop {} {\emph {\bibinfo {booktitle} {Advances in Artificial Life: 7th European Conference, ECAL 2003, Dortmund, Germany, September 14-17, 2003. Proceedings 7}}}\ (\bibinfo {organization} {Springer},\ \bibinfo {year} {2003})\ pp.\ \bibinfo {pages} {588--597}\BibitemShut {NoStop}%
\bibitem [{\citenamefont {Maksymov}\ and\ \citenamefont {Pototsky}(2023)}]{maksymov2023reservoir}%
  \BibitemOpen
  \bibfield  {author} {\bibinfo {author} {\bibfnamefont {I.~S.}\ \bibnamefont {Maksymov}}\ and\ \bibinfo {author} {\bibfnamefont {A.}~\bibnamefont {Pototsky}},\ }\bibfield  {title} {\bibinfo {title} {Reservoir computing based on solitary-like waves dynamics of liquid film flows: A proof of concept},\ }\href@noop {} {\bibfield  {journal} {\bibinfo  {journal} {Europhysics Letters}\ }\textbf {\bibinfo {volume} {142}},\ \bibinfo {pages} {43001} (\bibinfo {year} {2023})}\BibitemShut {NoStop}%
\bibitem [{\citenamefont {Nakajima}(2020)}]{nakajima2020physical}%
  \BibitemOpen
  \bibfield  {author} {\bibinfo {author} {\bibfnamefont {K.}~\bibnamefont {Nakajima}},\ }\bibfield  {title} {\bibinfo {title} {Physical reservoir computing—an introductory perspective},\ }\href@noop {} {\bibfield  {journal} {\bibinfo  {journal} {Japanese Journal of Applied Physics}\ }\textbf {\bibinfo {volume} {59}},\ \bibinfo {pages} {060501} (\bibinfo {year} {2020})}\BibitemShut {NoStop}%
\bibitem [{\citenamefont {Marcucci}\ \emph {et~al.}(2023)\citenamefont {Marcucci}, \citenamefont {Caramazza},\ and\ \citenamefont {Shrivastava}}]{marcucci2023new}%
  \BibitemOpen
  \bibfield  {author} {\bibinfo {author} {\bibfnamefont {G.}~\bibnamefont {Marcucci}}, \bibinfo {author} {\bibfnamefont {P.}~\bibnamefont {Caramazza}},\ and\ \bibinfo {author} {\bibfnamefont {S.}~\bibnamefont {Shrivastava}},\ }\bibfield  {title} {\bibinfo {title} {A new paradigm of reservoir computing exploiting hydrodynamics},\ }\href@noop {} {\bibfield  {journal} {\bibinfo  {journal} {arXiv preprint arXiv:2302.01978}\ } (\bibinfo {year} {2023})}\BibitemShut {NoStop}%
\bibitem [{\citenamefont {Marcucci}\ \emph {et~al.}(2020)\citenamefont {Marcucci}, \citenamefont {Pierangeli},\ and\ \citenamefont {Conti}}]{marcucci2020theory}%
  \BibitemOpen
  \bibfield  {author} {\bibinfo {author} {\bibfnamefont {G.}~\bibnamefont {Marcucci}}, \bibinfo {author} {\bibfnamefont {D.}~\bibnamefont {Pierangeli}},\ and\ \bibinfo {author} {\bibfnamefont {C.}~\bibnamefont {Conti}},\ }\bibfield  {title} {\bibinfo {title} {Theory of neuromorphic computing by waves: machine learning by rogue waves, dispersive shocks, and solitons},\ }\href@noop {} {\bibfield  {journal} {\bibinfo  {journal} {Physical Review Letters}\ }\textbf {\bibinfo {volume} {125}},\ \bibinfo {pages} {093901} (\bibinfo {year} {2020})}\BibitemShut {NoStop}%
\bibitem [{\citenamefont {Mussel}\ and\ \citenamefont {Schneider}(2019{\natexlab{a}})}]{mussel2019similarities}%
  \BibitemOpen
  \bibfield  {author} {\bibinfo {author} {\bibfnamefont {M.}~\bibnamefont {Mussel}}\ and\ \bibinfo {author} {\bibfnamefont {M.~F.}\ \bibnamefont {Schneider}},\ }\bibfield  {title} {\bibinfo {title} {Similarities between action potentials and acoustic pulses in a van der waals fluid},\ }\href@noop {} {\bibfield  {journal} {\bibinfo  {journal} {Scientific Reports}\ }\textbf {\bibinfo {volume} {9}},\ \bibinfo {pages} {2467} (\bibinfo {year} {2019}{\natexlab{a}})}\BibitemShut {NoStop}%
\bibitem [{\citenamefont {Mussel}\ and\ \citenamefont {Schneider}(2019{\natexlab{b}})}]{mussel2019sounds}%
  \BibitemOpen
  \bibfield  {author} {\bibinfo {author} {\bibfnamefont {M.}~\bibnamefont {Mussel}}\ and\ \bibinfo {author} {\bibfnamefont {M.~F.}\ \bibnamefont {Schneider}},\ }\bibfield  {title} {\bibinfo {title} {It sounds like an action potential: unification of electrical, chemical and mechanical aspects of acoustic pulses in lipids},\ }\href@noop {} {\bibfield  {journal} {\bibinfo  {journal} {Journal of the Royal Society Interface}\ }\textbf {\bibinfo {volume} {16}},\ \bibinfo {pages} {20180743} (\bibinfo {year} {2019}{\natexlab{b}})}\BibitemShut {NoStop}%
\bibitem [{\citenamefont {Severa}\ \emph {et~al.}(2016)\citenamefont {Severa}, \citenamefont {Parekh}, \citenamefont {Carlson}, \citenamefont {James},\ and\ \citenamefont {Aimone}}]{severa2016spiking}%
  \BibitemOpen
  \bibfield  {author} {\bibinfo {author} {\bibfnamefont {W.}~\bibnamefont {Severa}}, \bibinfo {author} {\bibfnamefont {O.}~\bibnamefont {Parekh}}, \bibinfo {author} {\bibfnamefont {K.~D.}\ \bibnamefont {Carlson}}, \bibinfo {author} {\bibfnamefont {C.~D.}\ \bibnamefont {James}},\ and\ \bibinfo {author} {\bibfnamefont {J.~B.}\ \bibnamefont {Aimone}},\ }\bibfield  {title} {\bibinfo {title} {Spiking network algorithms for scientific computing},\ }in\ \href@noop {} {\emph {\bibinfo {booktitle} {2016 IEEE international conference on rebooting computing (ICRC)}}}\ (\bibinfo {organization} {IEEE},\ \bibinfo {year} {2016})\ pp.\ \bibinfo {pages} {1--8}\BibitemShut {NoStop}%
\bibitem [{\citenamefont {Heyder}\ and\ \citenamefont {Schumacher}(2021)}]{heyder2021echo}%
  \BibitemOpen
  \bibfield  {author} {\bibinfo {author} {\bibfnamefont {F.}~\bibnamefont {Heyder}}\ and\ \bibinfo {author} {\bibfnamefont {J.}~\bibnamefont {Schumacher}},\ }\bibfield  {title} {\bibinfo {title} {Echo state network for two-dimensional turbulent moist rayleigh-b{\'e}nard convection},\ }\href@noop {} {\bibfield  {journal} {\bibinfo  {journal} {Physical Review E}\ }\textbf {\bibinfo {volume} {103}},\ \bibinfo {pages} {053107} (\bibinfo {year} {2021})}\BibitemShut {NoStop}%
\bibitem [{\citenamefont {Siegelmann}(1995)}]{siegelmann1995computation}%
  \BibitemOpen
  \bibfield  {author} {\bibinfo {author} {\bibfnamefont {H.~T.}\ \bibnamefont {Siegelmann}},\ }\bibfield  {title} {\bibinfo {title} {Computation beyond the turing limit},\ }\href@noop {} {\bibfield  {journal} {\bibinfo  {journal} {Science}\ }\textbf {\bibinfo {volume} {268}},\ \bibinfo {pages} {545} (\bibinfo {year} {1995})}\BibitemShut {NoStop}%
\bibitem [{\citenamefont {Noy}\ and\ \citenamefont {Darling}(2023)}]{noy2023nanofluidic}%
  \BibitemOpen
  \bibfield  {author} {\bibinfo {author} {\bibfnamefont {A.}~\bibnamefont {Noy}}\ and\ \bibinfo {author} {\bibfnamefont {S.~B.}\ \bibnamefont {Darling}},\ }\bibfield  {title} {\bibinfo {title} {Nanofluidic computing makes a splash},\ }\href@noop {} {\bibfield  {journal} {\bibinfo  {journal} {Science}\ }\textbf {\bibinfo {volume} {379}},\ \bibinfo {pages} {143} (\bibinfo {year} {2023})}\BibitemShut {NoStop}%
\bibitem [{\citenamefont {Sharma}\ and\ \citenamefont {Marcucci}(2022)}]{sharma2022navier}%
  \BibitemOpen
  \bibfield  {author} {\bibinfo {author} {\bibfnamefont {S.}~\bibnamefont {Sharma}}\ and\ \bibinfo {author} {\bibfnamefont {G.}~\bibnamefont {Marcucci}},\ }\bibfield  {title} {\bibinfo {title} {From {N}avier-{S}tokes millennium-prize problem to soft matter computing},\ }\href@noop {} {\bibfield  {journal} {\bibinfo  {journal} {arXiv preprint arXiv:2212.01492}\ } (\bibinfo {year} {2022})}\BibitemShut {NoStop}%
\bibitem [{\citenamefont {Sharma}\ \emph {et~al.}(2022{\natexlab{a}})\citenamefont {Sharma}, \citenamefont {Marcucci},\ and\ \citenamefont {Mahmud}}]{sharma2022complexity}%
  \BibitemOpen
  \bibfield  {author} {\bibinfo {author} {\bibfnamefont {S.}~\bibnamefont {Sharma}}, \bibinfo {author} {\bibfnamefont {G.}~\bibnamefont {Marcucci}},\ and\ \bibinfo {author} {\bibfnamefont {A.}~\bibnamefont {Mahmud}},\ }\bibfield  {title} {\bibinfo {title} {A complexity perspective on fluid mechanics},\ }\href@noop {} {\bibfield  {journal} {\bibinfo  {journal} {arXiv preprint arXiv:2212.00153}\ } (\bibinfo {year} {2022}{\natexlab{a}})}\BibitemShut {NoStop}%
\bibitem [{\citenamefont {Sharma}\ \emph {et~al.}(2024)\citenamefont {Sharma}, \citenamefont {Mahmud}, \citenamefont {Tarabella}, \citenamefont {Mougoyannis},\ and\ \citenamefont {Adamatzky}}]{sharma2024information}%
  \BibitemOpen
  \bibfield  {author} {\bibinfo {author} {\bibfnamefont {S.}~\bibnamefont {Sharma}}, \bibinfo {author} {\bibfnamefont {A.}~\bibnamefont {Mahmud}}, \bibinfo {author} {\bibfnamefont {G.}~\bibnamefont {Tarabella}}, \bibinfo {author} {\bibfnamefont {P.}~\bibnamefont {Mougoyannis}},\ and\ \bibinfo {author} {\bibfnamefont {A.}~\bibnamefont {Adamatzky}},\ }\bibfield  {title} {\bibinfo {title} {Information-theoretic language of proteinoid gels: Boolean gates and qr codes},\ }\href@noop {} {\bibfield  {journal} {\bibinfo  {journal} {arXiv preprint arXiv:2405.19337}\ } (\bibinfo {year} {2024})}\BibitemShut {NoStop}%
\bibitem [{\citenamefont {Giorgio}\ \emph {et~al.}(2025)\citenamefont {Giorgio}, \citenamefont {D'Angelo}, \citenamefont {Tarabella},\ and\ \citenamefont {Sharma}}]{degiorgio2025}%
  \BibitemOpen
  \bibfield  {author} {\bibinfo {author} {\bibfnamefont {G.~D.}\ \bibnamefont {Giorgio}}, \bibinfo {author} {\bibfnamefont {P.}~\bibnamefont {D'Angelo}}, \bibinfo {author} {\bibfnamefont {G.}~\bibnamefont {Tarabella}},\ and\ \bibinfo {author} {\bibfnamefont {S.}~\bibnamefont {Sharma}},\ }\href {https://arxiv.org/abs/2504.11401} {\bibinfo {title} {Advances in prebiotic chemistry: the potential of analog computing and navier-stokes nernst-planck (npns) modeling in organic electronics technologies (oects)}} (\bibinfo {year} {2025}),\ \Eprint {https://arxiv.org/abs/2504.11401} {arXiv:2504.11401 [physics.flu-dyn]} \BibitemShut {NoStop}%
\bibitem [{\citenamefont {Pitingolo}\ \emph {et~al.}(2018)\citenamefont {Pitingolo}, \citenamefont {Nizard}, \citenamefont {Riaud},\ and\ \citenamefont {Taly}}]{pitingolo2018beyond}%
  \BibitemOpen
  \bibfield  {author} {\bibinfo {author} {\bibfnamefont {G.}~\bibnamefont {Pitingolo}}, \bibinfo {author} {\bibfnamefont {P.}~\bibnamefont {Nizard}}, \bibinfo {author} {\bibfnamefont {A.}~\bibnamefont {Riaud}},\ and\ \bibinfo {author} {\bibfnamefont {V.}~\bibnamefont {Taly}},\ }\bibfield  {title} {\bibinfo {title} {Beyond the on/off chip trade-off: A reversibly sealed microfluidic platform for 3d tumor microtissue analysis},\ }\href@noop {} {\bibfield  {journal} {\bibinfo  {journal} {Sensors and Actuators B: Chemical}\ }\textbf {\bibinfo {volume} {274}},\ \bibinfo {pages} {393} (\bibinfo {year} {2018})}\BibitemShut {NoStop}%
\bibitem [{\citenamefont {Dhar}\ \emph {et~al.}(2020)\citenamefont {Dhar}, \citenamefont {Narendren}, \citenamefont {Gaur}, \citenamefont {Sharma}, \citenamefont {Kumar},\ and\ \citenamefont {Katiyar}}]{dhar2020self}%
  \BibitemOpen
  \bibfield  {author} {\bibinfo {author} {\bibfnamefont {P.}~\bibnamefont {Dhar}}, \bibinfo {author} {\bibfnamefont {S.}~\bibnamefont {Narendren}}, \bibinfo {author} {\bibfnamefont {S.~S.}\ \bibnamefont {Gaur}}, \bibinfo {author} {\bibfnamefont {S.}~\bibnamefont {Sharma}}, \bibinfo {author} {\bibfnamefont {A.}~\bibnamefont {Kumar}},\ and\ \bibinfo {author} {\bibfnamefont {V.}~\bibnamefont {Katiyar}},\ }\bibfield  {title} {\bibinfo {title} {Self-propelled cellulose nanocrystal based catalytic nanomotors for targeted hyperthermia and pollutant remediation applications},\ }\href@noop {} {\bibfield  {journal} {\bibinfo  {journal} {International journal of biological macromolecules}\ }\textbf {\bibinfo {volume} {158}},\ \bibinfo {pages} {1020} (\bibinfo {year} {2020})}\BibitemShut {NoStop}%
\bibitem [{\citenamefont {Sharma}\ \emph {et~al.}(2023)\citenamefont {Sharma}, \citenamefont {Mahmud}, \citenamefont {Tarabella}, \citenamefont {Mougoyannis},\ and\ \citenamefont {Adamatzky}}]{sharma2023morphological}%
  \BibitemOpen
  \bibfield  {author} {\bibinfo {author} {\bibfnamefont {S.}~\bibnamefont {Sharma}}, \bibinfo {author} {\bibfnamefont {A.}~\bibnamefont {Mahmud}}, \bibinfo {author} {\bibfnamefont {G.}~\bibnamefont {Tarabella}}, \bibinfo {author} {\bibfnamefont {P.}~\bibnamefont {Mougoyannis}},\ and\ \bibinfo {author} {\bibfnamefont {A.}~\bibnamefont {Adamatzky}},\ }\bibfield  {title} {\bibinfo {title} {On morphological and functional complexity of proteinoid microspheres},\ }\href@noop {} {\bibfield  {journal} {\bibinfo  {journal} {arXiv preprint arXiv:2306.11458}\ } (\bibinfo {year} {2023})}\BibitemShut {NoStop}%
\bibitem [{\citenamefont {Sharma}\ \emph {et~al.}(2022{\natexlab{b}})\citenamefont {Sharma}, \citenamefont {Mougoyannis}, \citenamefont {Tarabella},\ and\ \citenamefont {Adamatzky}}]{sharma2022review}%
  \BibitemOpen
  \bibfield  {author} {\bibinfo {author} {\bibfnamefont {S.}~\bibnamefont {Sharma}}, \bibinfo {author} {\bibfnamefont {P.}~\bibnamefont {Mougoyannis}}, \bibinfo {author} {\bibfnamefont {G.}~\bibnamefont {Tarabella}},\ and\ \bibinfo {author} {\bibfnamefont {A.}~\bibnamefont {Adamatzky}},\ }\bibfield  {title} {\bibinfo {title} {A review on the protocols for the synthesis of proteinoids},\ }\href@noop {} {\bibfield  {journal} {\bibinfo  {journal} {arXiv preprint arXiv:2212.02261}\ } (\bibinfo {year} {2022}{\natexlab{b}})}\BibitemShut {NoStop}%
\bibitem [{Note1()}]{Note1}%
  \BibitemOpen
  \bibinfo {note} {The weights $w_i$ are as follows: Number of Nodes: 0.05, Number of Edges: 0.05, Average Degree: 0.10, Clustering Coefficient: 0.10, Graph Density: 0.20, Degree Entropy: 0.20, Number of Connected Components: 0.10, and Average Resistance: 0.20. These weights were chosen to balance the contribution of each metric, with slightly higher emphasis on degree entropy and average resistance as they capture more nuanced aspects of the graph structure.}\BibitemShut {Stop}%
\bibitem [{\citenamefont {Schmidt-Hieber}(2021)}]{schmidt2021kolmogorov}%
  \BibitemOpen
  \bibfield  {author} {\bibinfo {author} {\bibfnamefont {J.}~\bibnamefont {Schmidt-Hieber}},\ }\bibfield  {title} {\bibinfo {title} {The kolmogorov--arnold representation theorem revisited},\ }\href@noop {} {\bibfield  {journal} {\bibinfo  {journal} {Neural networks}\ }\textbf {\bibinfo {volume} {137}},\ \bibinfo {pages} {119} (\bibinfo {year} {2021})}\BibitemShut {NoStop}%
\end{thebibliography}%

\end{document}